\documentclass[fp,twocolumn]{jpsj3}
\usepackage{txfonts,color}

\title{Spin Dynamics in the $t$-$t'$-$J$ Model: Dynamical Density-Matrix Renormalization Group Study}

\author{Takami Tohyama$^1$\thanks{tohyama@rs.tus.ac.jp}, Shigetoshi Sota$^2$, and Seiji Yunoki$^{2,3,4}$}
\inst{$^1$Department of Applied Physics, Tokyo University of Science, Tokyo 125-8585, Japan \\
$^2$Computational Materials Science Research Team, RIKEN Center for Computational Science (R-CCS), Kobe, Hyogo 650-0047, Japan \\
$^3$Computational Condensed Matter Physics Laboratory, RIKEN Cluster for Pioneering Research (CPR), Wako, Saitama 351-0198, Japan \\
$^3$Computational Quantum Matter Research Team, RIKEN Center for Emergent Matter Science (CEMS), Wako, Saitama 351-0198, Japan} 

\abst{The ground state of a hole-doped $t$-$t'$-$J$ ladder with four legs favors a striped charge distribution. Spin excitation from the striped ground state is known to exhibit incommensurate spin excitation near $\mathbf{q}=(\pi,\pi)$ along the leg direction ($q_x$ direction). However, an outward dispersion from the incommensurate position toward $\mathbf{q}=(0,\pi)$ is strong in intensity, inconsistent with inelastic neutron scattering (INS) experiment in hole-doped cuprates. Motivated by this inconsistency, we use the $t$-$t'$-$J$ model with $m\times n=96$ lattice sites by changing lattice geometry from four-leg ($24\times 4$) to rectangle ($12\times 8$) shape and investigate the dynamical spin structure factor by using the dynamical density matrix renormalization group. We find that the outward dispersion has weak spectral weights in the $12\times 8$ lattice, accompanied with the decrease of excitation energy close to $\mathbf{q}=(\pi,\pi)$, being consistent with the INS data. In the $12\times 8$ lattice, weakening of incommensurate spin correlation is realized even in the presence of the striped charge distribution.  For further investigation of geometry related spin dynamics, we focus on direction dependent spin excitation reported by recent resonant inelastic x-ray scattering (RIXS) for cuprate superconductors and obtain a consistent result with RIXS by examining an $8\times 8$ $t$-$t'$-$J$ square lattice.}

\begin{document}
\maketitle

\section{Introduction}

In hole-doped cuprate superconductors, an hourglass-type spin excitation centered at  the magnetic zone center in the Brillouin zone has been observed by inelastic neutron scattering (INS) experiment~\cite{Fujita2012}. One of possible origins of the hourglass-type excitation is the formation of charge stripes in hole-doped cuprates~\cite{Tranquada1995} as discussed based on a two-dimensional (2D) single-band Hubbard model~\cite{Kaneshita2001,Seibold2005,Seibold2006} and a localized spin model~\cite{Kruger2003,Carlson2004}. Recent quantum Monte Carlo (QMC) calculations of the dynamical spin structure factor for a four-leg, three-band Hubbard ladder including oxygen orbitals~\cite{Huang2017} and for a $t$-$t'$-$U$ four-leg ladder~\cite{Huang2017b} have also indicated the hourglass-type excitation in the presence of the charge stripes where periodical arrangement of a river of charge on rung is formed along the leg direction. 

A hole-doped four-leg $t$-$t'$-$J$ ladder has also shown the charge-stripe ground state~\cite{Tohyama1999,White1999,Scalapino2012,Dodaroo2017} and clear incommensurate spin excitation near the magnetic zone center at $\mathbf{q}=(\pi,\pi)$ forming an hourglass behavior as demonstrated by using the dynamical version of the density-matrix renormalization group (DMRG)~\cite{Tohyama2018}. The hourglass behavior qualitatively agrees with the experimental data when one uses ladders with four legs, but the presence of an outward dispersion with strong spectral weight from the incommensurate position toward $\mathbf{q}=(0,\pi)$~\cite{Tohyama2018} is inconsistent with experimental observations~\cite{Fujita2012}.  The calculated energy of the $\mathbf{q}=(\pi,\pi)$ excitation is nearly the same as the value of antiferromagnetic exchange interaction $J$~\cite{Tohyama2018}, being also inconsistent with experimental observation showing less than half of $J$~\cite{Fujita2012}.  It is therefore crucially important to clarify the origin of these inconsistencies.

The number of hole carrier changes spin excitation in cuprates. Recent measurements of spin excitation in La$_{2-x}$Sr$_x$CuO$_4$ by resonant inelastic x-ray scattering (RIXS) tuned for the Cu $L$ edge have shown that energy difference between peak positions near $\mathbf{q}=(\pi,0)$ and $\mathbf{q}=(\pi/2,\pi/2)$ increases with increasing hole carriers~\cite{Meyers2017,Robarts2019}. In fact, the peak energy at $\mathbf{q}=(\pi/2,\pi/2)$ is almost a half of that at $\mathbf{q}=(\pi,0)$ in the overdoped region~\cite{Meyers2017,Robarts2019,Guarise2014,Monney2016}. It is unclear whether the $t$-$t'$-$J$ model can explain this behavior, though there are mean-field-type calculations based on random-phase approximation (RPA) for a $t$-$t'$-$U$ Hubbard model~\cite{Guarise2014,Monney2016} and QMC simulations for a three-band Hubbard model~\cite{Peng2018}. 

In this paper, we investigate dynamical spin structure factor in the $t$-$t'$-$J$ model  by using dynamical DMRG to deepen our understanding on the two issues mentioned above. For this purpose, we use $m\times n$ $t$-$t'$-$J$ lattices. In terms of incommensurate spin excitation near $\mathbf{q}=(\pi,\pi)$, we fix $m\times n=96$ and examine three lattice shapes: $m\times n=24\times 4$, $16\times 6$, and $12\times 8$. Changing geometry from ladder shape ($24\times 4$) to rectangle shape ($12\times 8$), we find that the outward dispersion from the incommensurate position looses its intensity and excitation energy very close to $\mathbf{q}=(\pi,\pi)$ decreases, resulting in spectral behavior consistent with INS data for cuprate superconductors~\cite{Fujita2012}.  These are accompanied with weakening of incommensurate spin correlation in the ground state even in the presence of striped charge distribution in the $12\times 8$ lattice. We also investigate the directional dependence of spin excitation along the $(0,0)$-$(\pi,0)$ and $(0,0)$-$(\pi,\pi)$ directions using an $8\times 8$ $t$-$t'$-$J$ square lattice. We find a different behavior along the two directions in terms of spin excitation energy at the overdoped region, qualitatively consistent with the RIXS data~\cite{Meyers2017,Robarts2019}, though softening of spin excitation with hole doping is stronger than the observed one by RIXS.

This paper is organized as follows. The $m\times n$ $t$-$t'$-$J$ lattices with fixed 96 sites ($24\times 4$, $16\times 6$, and $12\times 8$) and dynamical DMRG method are introduced  in Sec.~\ref{Sec2}. In Sec.~\ref{Sec3}, we calculate the dynamical spin and charge structure factors near $\mathbf{q}=(\pi,\pi)$ and make clear the dependence of the spin structure factor on lattice geometry. The doping dependence of spin excitation for the $12\times 8$ lattice near $\mathbf{q}=(\pi,\pi)$ is also shown. In Sec.~\ref{Sec4}, the directional dependence of the dynamical spin structure factor is examined by using an $8\times 8$ square lattice. Finally, a summary is given in Sec.~\ref{Sec5}.

\section{Model and method}
\label{Sec2}
The Hamiltonian of the hole-doped $t$-$t'$-$J$ model in two dimensions reads 
\begin{eqnarray}
H&=& -t\sum_{\mathbf{l},\boldsymbol{\delta},\sigma}
    \left( \tilde{c}_{\mathbf{l}+\boldsymbol{\delta},\sigma }^\dagger \tilde{c}_{\mathbf{l},\sigma}+\tilde{c}_{\mathbf{l}-\boldsymbol{\delta},\sigma }^\dagger \tilde{c}_{\mathbf{l},\sigma} \right) \nonumber \\
&&  -t'\sum_{\mathbf{l},\boldsymbol{\delta}',\sigma}
   \left( \tilde{c}_{\mathbf{l}+\boldsymbol{\delta}',\sigma }^\dagger \tilde{c}_{\mathbf{l},\sigma }+\tilde{c}_{\mathbf{l}-\boldsymbol{\delta}',\sigma }^\dagger \tilde{c}_{\mathbf{l},\sigma } \right) \nonumber \\
&&  +J\sum_{\mathbf{l},\boldsymbol{\delta}}
      \left( \mathbf{S}_{\mathbf{l}+\boldsymbol{\delta}}\cdot \mathbf{S}_\mathbf{l}-\frac{1}{4} n_{\mathbf{l}+\boldsymbol{\delta}} n_{\mathbf{l}}\right) ,
\label{H}
\end{eqnarray}
where $t$, $t'$, and $J$ are the nearest-neighbor hopping, the next-nearest-neighbor hopping, and the antiferromagnetic exchange interaction, respectively; $\boldsymbol{\delta}=\mathbf{x}$, $\mathbf{y}$ and $\boldsymbol{\delta}'=\mathbf{x}+\mathbf{y}$, $\mathbf{x}-\mathbf{y}$, with $\mathbf{x}$ and $\mathbf{y}$ being the unit vectors in the $x$ and $y$ directions, respectively; the operator $\tilde{c}_{\mathbf{l},\sigma}=c_{\mathbf{l},\sigma}(1-n_{\mathbf{l},-\sigma})$, with $n_{\mathbf{l},\sigma}=c_{\mathbf{l},\sigma}^\dagger c_{\mathbf{l},\sigma}$, annihilates a localized electron with spin $\sigma$ at site $\mathbf{l}$ with the constraint of no double occupancy; $\mathbf{S}_\mathbf{l}$ is the spin operator at site $\mathbf{l}$; and $n_\mathbf{l}=n_{\mathbf{l},\uparrow}+n_{\mathbf{l},\downarrow}$. In the following calculations, we fix $J/t=0.4$ and $t'/t=-0.25$, which are typical values appropriate for cuprates with $t\sim0.35$~eV.

We use $m\times n=96$-site lattices with cylindrical geometry, where the $x$ direction with $m$ sites has an open boundary condition while the $y$ direction with $n$ sites has a periodic boundary condition. We consider three cases: $(m,n)=(24,4)$, $(16,6)$, and $(12,8)$. The hole density for $N_\mathrm{h}$ holes in the lattices is defined by $x=N_\mathrm{h}/96$.  In the $m\times n$ lattice, the $y$ component of momentum $\mathbf{q}$ is determined by using standard translational symmetry, i.e., $q_y=2n_y\pi/n$ ($n_y=0, \pm 1,\cdots,\pm(n/2-1), n/2$), but the $x$ component is given by $q_x=n_x\pi/(m+1)$ ($n_x=1,2,\cdots,m$) because of the open boundary condition. Defining $l_x$ ($l_y$) as the $x$ ($y$) component of site $\mathbf{l}$, we can write the Fourier component for the $z$ component of spin operator and that of charge operator as 
\begin{equation}
S_\mathbf{q}^z=\sqrt{\frac{2}{(m+1)n}} \sum_\mathbf{l} \sin(q_x l_x) e^{-iq_y l_y}S_\mathbf{l}^z \;,
\label{Sz}
\end{equation}
and
\begin{equation}
N_\mathbf{q}=\sqrt{\frac{2}{(m+1)n}} \sum_\mathbf{l} \sin(q_x l_x) e^{-iq_y l_y}n_\mathbf{l} \;,
\label{N}
\end{equation}
respectively. 

The dynamical spin and charge structure factors, $S(\mathbf{q},\omega)$ and $N(\mathbf{q},\omega)$, are defined as
\begin{equation}
S(\mathbf{q},\omega)=-\frac{1}{\pi} \mathrm{Im} \left\langle 0 \right| \tilde{S}_{-\mathbf{q}}^z \frac{1}{\omega  - H + E_0+i\gamma } \tilde{S}_\mathbf{q}^z \left| 0 \right\rangle \label{Sqw} \;,
\end{equation}
and
\begin{equation}
N(\mathbf{q},\omega)=-\frac{1}{\pi} \mathrm{Im} \left\langle 0 \right| \tilde{N}_{-\mathbf{q}} \frac{1}{\omega  - H + E_0+i\gamma } \tilde{N}_\mathbf{q} \left| 0 \right\rangle \;, \label{Nqw}
\end{equation}
where $\left|0 \right\rangle$ represents the ground state with energy $E_0$, $\tilde{S}_\mathbf{q}=S^z_\mathbf{q}-\left\langle 0\right| S^z_\mathbf{q} \left| 0\right\rangle$, $\tilde{N}_\mathbf{q}=N_\mathbf{q}-\left\langle 0\right| N_\mathbf{q} \left| 0\right\rangle$, and $\gamma$ is a small positive number.
The static spin structure factor is defined as
$S(\mathbf{q})=\left\langle 0 \right| \tilde{S}_{-\mathbf{q}}^z \tilde{S}_\mathbf{q}^z \left| 0 \right\rangle$.

We calculate Eqs.~(\ref{Sqw}) and (\ref{Nqw}) for the $m\times n$ $t$-$t'$-$J$ lattice using dynamical DMRG, where we use three kinds of target states: for $S(\mathbf{q},\omega)$, (i) $\left| 0\right\rangle$, (ii) $S^z_\mathbf{q} \left| 0 \right\rangle$, and (iii) $(\omega - H + E_0+i\gamma)^{-1} S^z_\mathbf{q} \left| 0 \right\rangle$, and for $N(\mathbf{q},\omega)$ we use $\tilde{N}_\mathbf{q}$ instead of $S^z_\mathbf{q}$. Target state (iii) is evaluated using a kernel-polynomial expansion method~\cite{Sota2010}, where the broadening of spectra becomes a Gaussian type with a half width at half maximum of $0.08t$. In our numerical calculations of $S(\mathbf{q},\omega)$, we divide the energy interval $[0,t]$ by 50 mesh points and target all of the points at once. 

To perform DMRG for the $m\times n$ $t$-$t'$-$J$ lattice, we construct a snakelike one-dimensional chain, which runs from the site $(1,1)$ to $(1,n)$, then from $(2,n)$ to $(2,1)$, and repeats this pattern until we reach the site $(m,1)$. We use the maximum truncation number $M=6000$. Resulting truncation error is less than $2.3\times10^{-4}$ for the ground state $\left| 0 \right\rangle$. To check convergence of our numerical results in terms of $M$, we performed dynamical DMRG calculations with $M=4000$ for $n=8$ and found that, for small $x$ ($x=8/96=0.083$), the deviation of the energy position and intensity for the highest peaks from those for $M=6000$ is within 5\%, showing a good convergence. For large $x$ ($x=16/96=0.167$), the deviation of peak position and intensity is maximally 10\% and 20\%, respectively, being less convergence. At present, the $M=6000$ calculation for the $m\times n$ $t$-$t'$-$J$ ladder is the best one that we can perform by our present computer resources. More time consuming calculations more than $M=6000$ remains as a future problem.

For square geometry discussed in Sec.~\ref{Sec4}, we use an $8\times 8$ $t$-$t'$-$J$ lattice with open boundary condition in both directions. In this lattice, we take $M=8000$, which is larger than $M=6000$ for the 96-site lattices but gives similar values of the truncation error less than $1.8\times 10^{-4}$ for the ground state. Therefore, the $8\times 8$ lattice is the maximum size with the square geometry that we can treat for our calculation of $S(\mathbf{q},\omega)$ at present.

\section{Geometry and doping dependence of the $t$-$t'$-$J$ model}
\label{Sec3}
\subsection{Geometry dependence of $S(\mathbf{q},\omega)$}
\label{Sec3.1}

\begin{figure}[tb]
\center{
\includegraphics[width=0.45\textwidth]{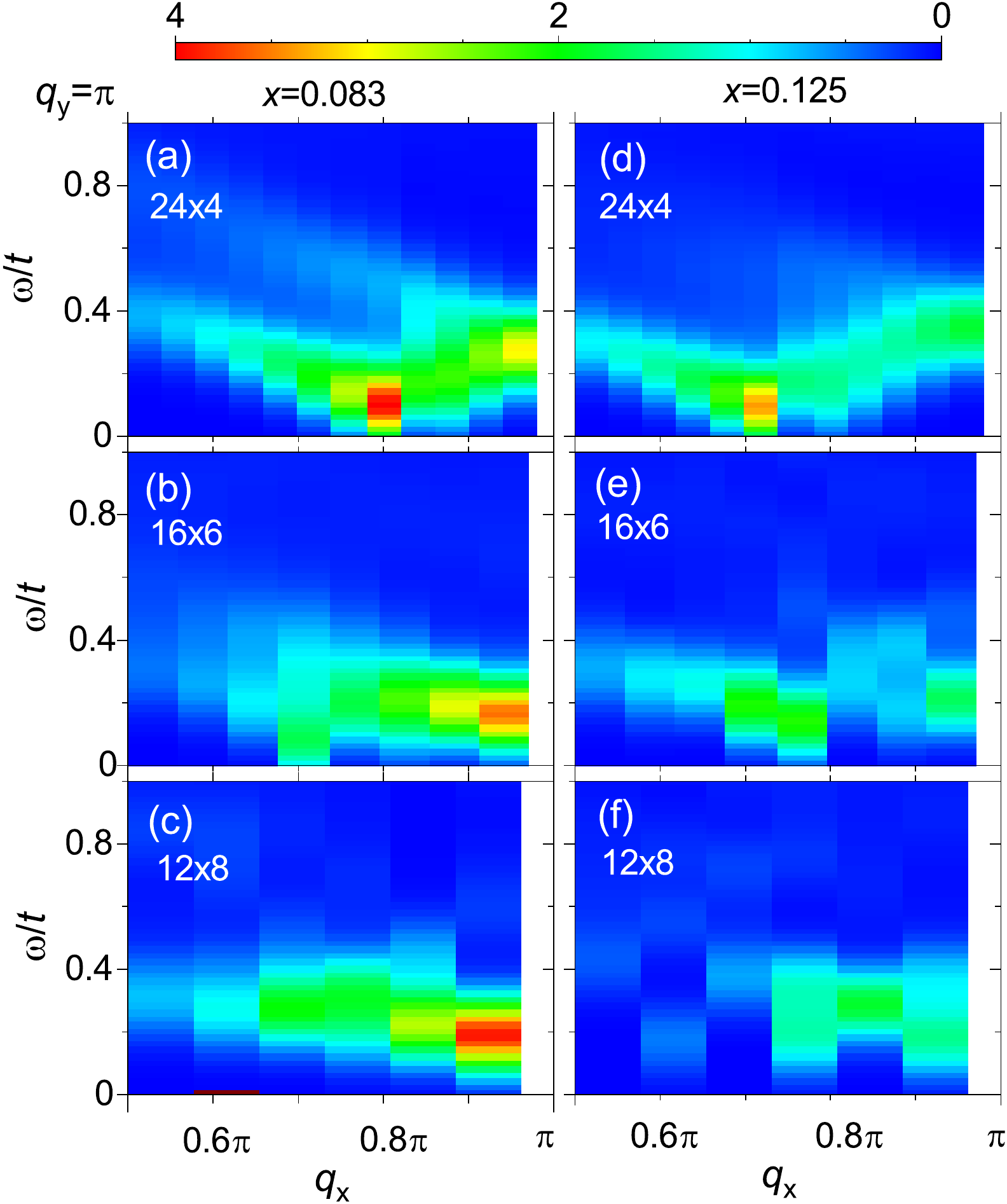}
}
\caption{(Color online) $S(\mathbf{q},\omega)$ from $\mathbf{q}=(0.5\pi,\pi)$ to $(\pi,\pi)$ in the $m\times n$ $t$-$t'$-$J$ lattice with $J/t=0.4$ and $t'/t=-0.25$. (a) $24\times 4$~\cite{Tohyama2018}, (b) $16\times 6$, and (c) $12\times 8$ for $x=0.083$. (d) $24\times 4$~\cite{Tohyama2018}, (e) $16\times 6$, and (f) $12\times 8$ for $x=0.125$.}
\label{fig1}
\end{figure}

We examine the effect of the geometry of lattice on $S(\mathbf{q},\omega)$ in the $m\times n=96$ $t$-$t'$-$J$ lattice. Figure~\ref{fig1} shows $S(\mathbf{q},\omega)$ from $\mathbf{q}=(0.5\pi,\pi)$ to $(\pi,\pi)$ for $x=8/96=0.083$ (left panels) and  $x=12/96=0.125$ (right panels). For the $24\times 4$ lattice~\cite{Tohyama2018}, there is a low-energy excitation at $q_\mathrm{IC}= \pi(1-2x)$ as shown in Figs.~\ref{fig1}(a) and \ref{fig1}(d), which is consistent with incommensurate vectors reported in hole-doped cuprate superconductor La$_{2-x}$Sr$_x$CuO$_4$~\cite{Yamada1998}. This low-energy excitation is originated from the formation of stripe order in the ground state~\cite{Tohyama2018}. Linear dispersive branches emerge from $q_\mathrm{IC}$ toward both the $q_x=\pi$ (inward) and $q_x=0$ (outward) directions. In the INS experiment~\cite{Fujita2012}, the outward dispersion has not been observed. Furthermore, in other calculations of $S(\mathbf{q},\omega)$ under the stripe order for the 2D extended Hubbard model based on RPA~\cite{Kaneshita2001} and time-dependent Gutzwiller approximation~\cite{Seibold2006}, the outward dispersion loses its intensity quickly for small $x$. Therefore, the present inconsistency may arise from ladder geometry with four legs in the $24\times 4$ lattice. In fact, in the $16\times 6$ lattice [Figs.~\ref{fig1}(b) and \ref{fig1}(e)] and the $12\times 8$ lattice [Figs.~\ref{fig1}(c) and \ref{fig1}(f)], spectral shape changes significantly. At $x=0.083$, spectral weight moves toward $q_x=\pi$ with approaching square geometry and incommensurate low-energy excitation disappears in the $12\times 8$ lattice. At $x=0.125$, the dispersive excitation in Fig.~\ref{fig1}(d) changes significantly with again approaching square geometry as seen in Figs.~\ref{fig1}(e) and \ref{fig1}(f). In the $12\times 8$ lattice, the spectral weight for $q_x<q_\mathrm{IC}$ becomes small and thus the outward dispersion cannot be seen in Fig.~\ref{fig1}(f), giving rise to a consistent behavior with the experimental observation.

The energy position of spin excitation close to $\mathbf{q}=(\pi,\pi)$ is also dependent on the geometry of lattice. At $x=0.125$, the corresponding position for the $24\times 4$ lattice is $\omega/t\sim 0.36$ close to $J/t=0.4$ [see the maximum intensity position at $q_x=0.96$ in Fig.~\ref{fig1}(d)]. The position shifts to the low-energy side with approaching square geometry, i.e., $\omega/t\sim 0.22$ for the $16\times 6$ lattice [see $q_x=0.94$ in Fig.~\ref{fig1}(e)] and $\omega/t\sim 0.18$ for the $12\times 8$ lattice [see $q_x=0.92$ in Fig.~\ref{fig1}(f)]. The energy $\omega/t\sim 0.2=J/2$ is quantitatively consistent with experimental observation~\cite{Fujita2012}. Therefore, it would be fair to say that the $12\times 8$ result at $x=0.125$ reasonably reproduces experimental behaviors of spin excitation around $\mathbf{q}=(\pi,\pi)$ observed by INS. 

It is interesting to note that spectral distribution at $q_x=0.85\pi$ for the $12\times 8$ lattice shown in Fig.~\ref{fig1}(f) is slightly higher in energy than that at neighboring $q_x$. This is similar to the case of the $16\times 6$ lattice but different from the case of the $24\times 4$ lattice shown in Fig.~\ref{fig1}(d), where the energy at $q_x=0.85\pi$ is the middle of those at $q_x\sim q_\mathrm{IC}$ and $q_x\sim\pi$. Therefore, the spectral distribution for the $12\times 8$ lattice at $x=0.125$ indicates that two contributions at $q_x\sim q_\mathrm{IC}$ and $q_x\sim\pi$ have different origin in contrast to the case of stripe-driven spin excitations in the four-leg $24\times 4$ lattice. This supports the idea that there is an outward dispersive excitation starting from the $(\pi,\pi)$ excitation, i.e., an upper part of hourglass spin excitation, which is independent of the incommensurate low-energy excitation~\cite{Sato2014,Fujita_private}. We note that spectral distribution in Fig.~\ref{fig1}(f) is different from an hourglass type in the sense that there is no smooth upward connection from $q_x\sim q_\mathrm{IC}$ to $q_x\sim\pi$.

In Fig~\ref{fig1}(f), we changed $q_x$ from $q_x=12/13\pi\sim 0.92\pi$. If we change $q_y$ from $\mathbf{q}=(0.92\pi,\pi)$, we can define only one momentum $q_y=0.75\pi$ toward $q_y=0.5\pi$. At the momentum $(0.92\pi,0.75\pi)$, we find a peak with energy $\omega/t\sim 0.17$ (not shown here), whose position is close to that of low-energy spectrum at $(0.77\pi,\pi)$ in Fig.~\ref{fig1}(f).

\begin{figure}[tb]
\center{
\includegraphics[width=0.45\textwidth]{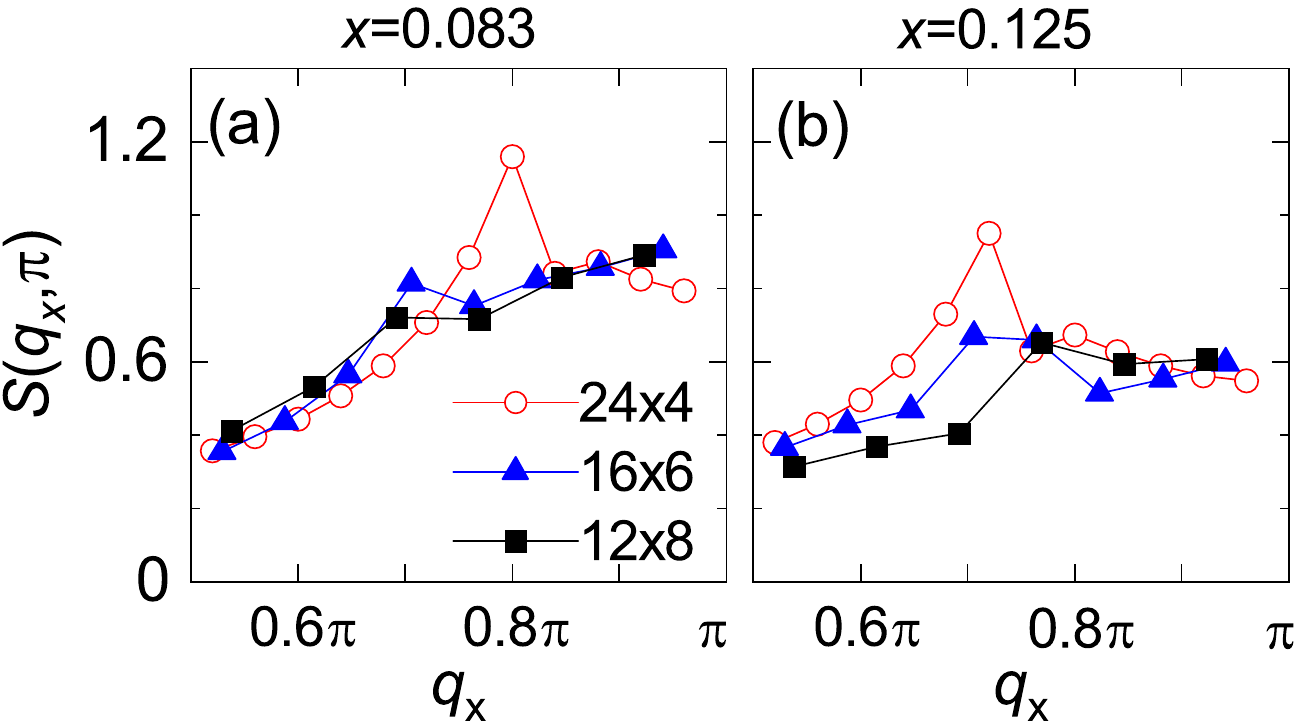}
}
\caption{(Color online) Static spin structure factor $S(\mathbf{q})$ for $\mathbf{q}=(q_x,\pi)$ for the $24\times 4$ (open read circles), $16\times 6$ (sold blue triangles), and $12\times 8$ (sold black squares) $t$-$t'$-$J$ lattice with $J/t=0.4$ and $t'/t=-0.25$. (a) $x=0.083$ and (b)  $x=0.125$.}
\label{fig2}
\end{figure}

To understand the change of spin excitation with changing lattice geometry, we show in Fig.~\ref{fig2} the static spin structure factor $S(\mathbf{q})$ along the same momentum direction as in the case of Fig.~\ref{fig1}, i.e., $\mathbf{q}=(q_x,\pi)$. In both cases of $x=0.083$ and $x=0.125$, the pronounced value of $S(\mathbf{q})$ near $q_\mathrm{IC}$ in the $24\times 4$ lattice decreases with changing lattice to $16\times 6$ and $12\times 8$. We also notice that, while $S(\mathbf{q})$ decreases toward $q_x=\pi$ from its maximum value near $q_x=q_\mathrm{IC}$ in the $24\times 4$ lattice, it increases toward $q_x=\pi$ in the $16\times 6$ and $12\times 8$ lattices. These contrasting behaviors indicate that, with changing geometry toward square lattice, incommensurate spin correlation weakens and local antiferromagnetic spin correlation becomes relatively strong in the systems. This seems to be consistent with the behavior that spin excitation near $\mathbf{q}=(\pi,\pi)$ is stronger in the $12\times 8$ lattice at both $x=0.083$ and $x=0.125$ as seen in Fig.~\ref{fig1}.

\begin{figure}[tb]
\center{
\includegraphics[width=0.45\textwidth]{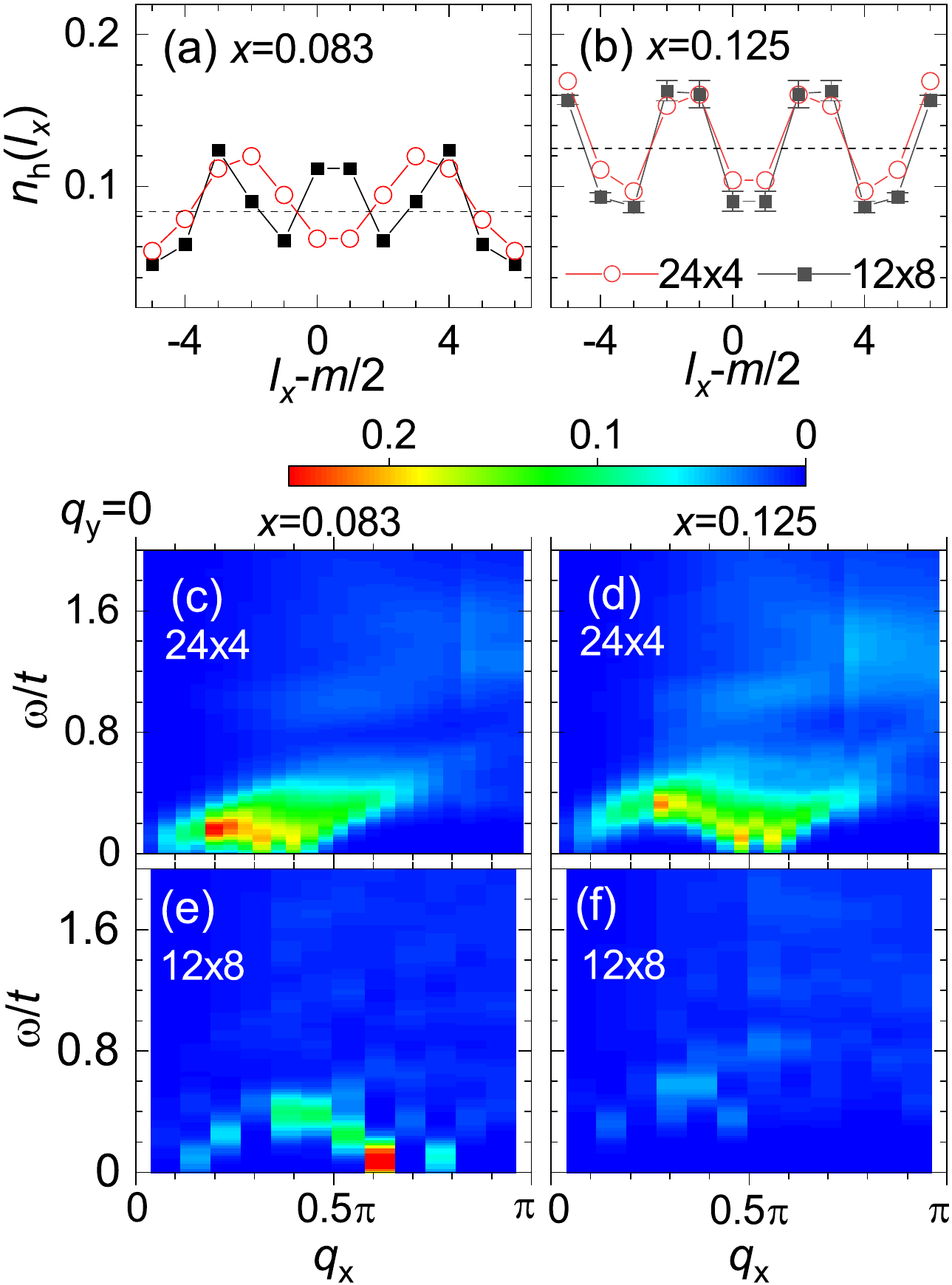}
}
\caption{(Color online) (a) Hole number $n_\mathrm{h}(l_x)$ at the leg position $l_x$ for $x=0.083$ and (b) that at $x=0.125$ in the $24\times 4$ ($m=24$ and red circles) and $12\times 8$ ($m=12$ and black squares) $t$-$t'$-$J$ lattice with $J/t=0.4$ and $t'/t=-0.25$. In (b), the bars attached to the black squares denote the minimum and maximum values of $n_\mathrm{h}(l_x)$ showing four-lattice periodicity along the $l_y$ direction. The horizontal dotted line denotes averaged hole density $x$. (c) and (d): $N(\mathbf{q},\omega)$ along $q_x$ with $q_y=0$ for the $24\times 4$ $t$-$t'$-$J$ lattice at $x=0.083$ and $x=0.125$~\cite{Tohyama2018}.  (e) and (f): the same as (c) and (d) but for the $12\times 8$ $t$-$t'$-$J$ lattice. The intensity above 0.24 at $q_x=0.62\pi$ in (e) is colored in red and the maximum intensity is 0.30 at $\omega/t=0.08$.}
\label{fig3}
\end{figure}

In the $24\times 4$ lattice, the hole number $n(l_x)$ along the leg position $l_x$ oscillates with period of $1/(2x)$ as a consequence of the formation of stripe charge order~\cite{White1999,Tohyama2018}. This stripe order accommodates incommensurate spin structure in the spin background, resulting in the $x$ dependent spin excitation seen in Figs.~\ref{fig1}(a) and \ref{fig1}(d)~\cite{Tohyama2018}. To clarify whether charge order plays a crucial role in spin excitation even in the $12\times 8$ lattice, we compare  $n(l_x)$ in the $12\times 8$ lattice with that in the $24\times 4$ lattice in Figs.~\ref{fig3}(a) and \ref{fig3}(b) for $x=0.083$ and $x=0.125$, respectively.  In Figs.~\ref{fig3}(c)-\ref{fig3}(f) we also compare $N(\mathbf{q},\omega)$ along the $q_x$ direction with $q_y=0$ in the $12\times 8$ lattice with that in the $24\times 4$ lattice for $x=0.083$ and $x=0.125$.

At $x=0.083$, $n(l_x)$ in the $12\times 8$ lattice exhibits an oscillation with amplitude similar to that in the $24\times 4$ lattice, but its oscillation period is roughly half [see Fig.~\ref{fig3}(a)]. This difference results in the difference of momentum for the lowest-energy charge excitation: $q_x\sim 0.35\pi$ for the $24\times 4$ lattice [Fig.~\ref{fig3}(c)] and $q_x\sim 0.6\pi$ for the $12\times 8$ lattice [Fig.~\ref{fig3}(e)].  However, we have to emphasize that there is no incommensurate spin excitation in the $12\times 8$ lattice as shown in Fig.~\ref{fig1}(c) in contrast to the case of the $24\times 4$ lattice [Fig.~\ref{fig1}(a)] where there is an incommensurate low-energy spin excitation at $q_x\sim 0.8\pi$ whose deviation from $q_x=\pi$, i.e., $\sim 0.2\pi$, is given by roughly a half of the momentum of stripe order $\sim 0.35\pi$. This means that the relation obtained in the $24\times 4$ lattice~\cite{Tohyama2018} cannot be applied for the $12\times 8$ lattice.

In contrast to the case of $x=0.083$, $n(l_x)$ at $x=0.125$ exhibits similar $l_x$ dependence between the $24\times 4$ and $12\times 8$ lattices as shown in Fig.~\ref{fig3}(b). The four-period oscillation along the $l_x$ direction is consistent with variational Monte Carlo result for the $t$-$t'$-$U$ Hubbard model~\cite{Ido2018}. We note that there is a small oscillation of hole number with four-lattice period along the $l_y$ direction for a given $l_x$, whose minimum and maximum values are shown by bars in Fig.~\ref{fig3}(b). Since there is transnational symmetry along the $l_y$ direction, this oscillation is due to insufficient number of $M$ in our calculation. Although a clear four-period oscillation is seen along the $l_x$ direction, $N(\mathbf{q},\omega)$ exhibits different behavior between Figs.~\ref{fig3}(d) and \ref{fig3}(f). For the $24\times 4$ lattice, a strong low-energy excitation appears at $q_x=0.5\pi$ equivalent to four-lattice periodicity. On the other hand, there is no such a strong low-energy excitation around $q_x=0.5\pi$ for the $12\times 8$ lattice, although low-energy spectral weight with energy $\omega/t=0.4$ exists at $q_x=0.46\pi$ [see Fig.~\ref{fig3}(f)] as a result of the static charge stripe with the four-period oscillation. We note that the spectral weight is distributed not only for $q_y=0$ but also for other $q_y$. This contrasting behavior in charge dynamics between the $24\times 4$ and $12\times 8$ lattices even in the presence of the same static charge order is in parallel with contrasting spin excitation discussed above. Namely, in the $24\times 4$ lattice the striped charge order induces spin correlation with double period of charge, while such a mechanism does not appear in the $12\times 8$ lattice and spin correlation remains short ranged even in the presence of the static charge stripe. The full understanding of microscopic origins on the contrasting behaviors remains as a future problem, but it may be interesting to investigate whether the spin dynamics in $12\times 8$ lattice is related to itinerant spin excitation due to quasi-one-dimensional Fermi surface~\cite{Yamase2001} since the charge stripe induces one-dimensional electronic states.

\subsection{Doping dependence of $S(\mathbf{q},\omega)$}
\label{Sec3.2}

\begin{figure}[tb]
\center{
\includegraphics[width=0.45\textwidth]{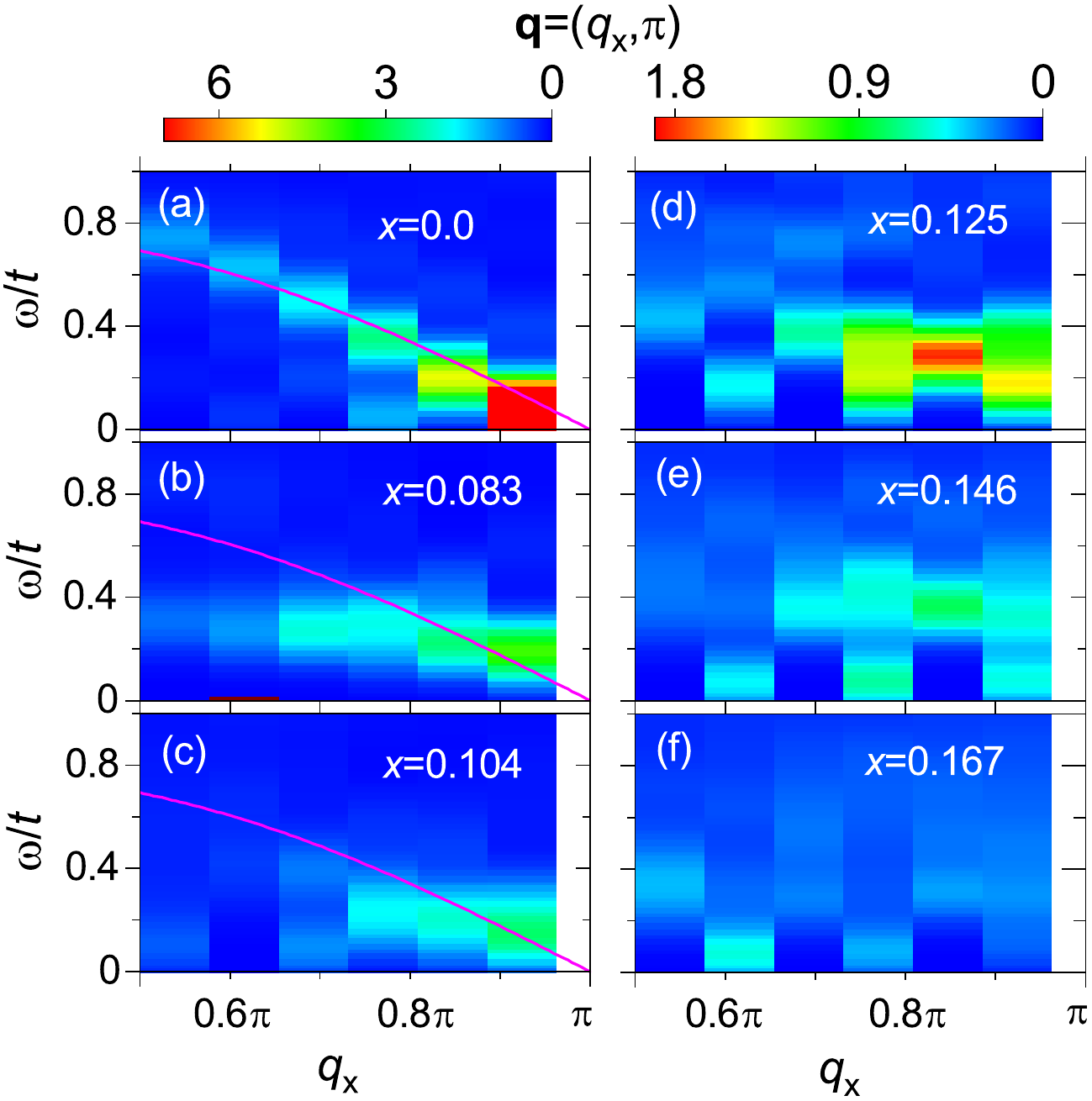}
}
\caption{(Color online) The $x$ dependence of $S(\mathbf{q},\omega)$ from $\mathbf{q}=(0.5\pi,\pi)$ to $(\pi,\pi)$ in the $12\times 8$ $t$-$t'$-$J$ ladder with $J/t=0.4$ and $t'/t=-0.25$. (a) $x=0$ (half filling), (b) $x=4/96=0.083$, (c) $x=8/96=0.104$, (d) $x=12/96=0.125$, (e) $14/96=0.146$, and (f) $x=16/96=0.167$. The intensity above 7 at $q_x=0.92\pi$ in (a) is colored in red and the maximum intensity is 19.8 at $\omega/t=0.06$. The purple line in (a)-(c) represents spin-wave dispersion at half filling obtained by the linear spin-wave theory for the 2D Heisenberg model.}
\label{fig4}
\end{figure}

Figure~\ref{fig4} shows the $x$ dependence of $S(\mathbf{q},\omega)$ from $\mathbf{q}=(0.5\pi,\pi)$ to $(\pi,\pi)$ in the $12\times 8$ $t$-$t'$-$J$ lattice. At $x=0$, dispersive spectral weight follows spin-wave dispersion as expected. We note that spectral weight at $q_x=0.92\pi$ and $0.85\pi$ is located slightly below the spin-wave dispersion and that there is a small low-energy weight below the spin-wave energy at $q_x=0.77\pi$, both of which are due to finite-size effect of the $12\times 8$ lattice with open boundary condition in the $x$ direction. In fact, there is a similar low-energy weight below the spin-wave energy around $q_x=0.7\pi$ even in the $16\times 6$ and $8\times 8$ lattices with open boundary condition (not shown here). 

With increasing $x$ from $x=0$ to $x=10/96=0.104$ shown in Fig.~\ref{fig4}(c), spectral weight at $q_x=0.92\pi$ decreases and low-energy spectral weight below the spin-wave dispersion spreads to smaller $q_x$ region. At $x=0.125$, low-energy excitation centered around $\omega/t=0.18$ at $q_x=0.77\pi$ appears as shown in Fig.~\ref{fig4}(d). Note that Fig.~\ref{fig4}(d) is the same figure as Fig.~\ref{fig1}(f) but their color scale is different. As discussed in Sec.~\ref{Sec3.1}, since spectral distribution at $q_x=0.85\pi$ is higher in energy than those at $q_x=0.77\pi$ $(\sim q_\mathrm{IC})$ and $q_x=0.92\pi$ $(\sim\pi)$ there is no smooth connection of spectral weight from $q_x=0.77\pi$ to $q_x=0.92\pi$, unlike the case for the $24\times 4$ lattice. This means that the spectral distribution for the $12\times 8$ lattice is different from an hourglass-type spectrum for the $24\times 4$ lattice. The present spectral distribution is rather consistent with the recently proposed view that an outward dispersive excitation starting from $\mathbf{q}=(\pi,\pi)$ and a dispersionless excitation starting from $\mathbf{q}=(q_\mathrm{IC},\pi)$ coexist~\cite{Sato2014,Fujita_private}.

With further increasing $x$, spectral weight is reduced and whole intensity tends to be distributed over all momentum region with small intensity. In fact, $S(q_x,\pi)$ corresponding to integrated spectral weight at $x=0.167$ shows small momentum dependence within $0.37<S(q_x,\pi)<0.42$ for $0.5\pi<q_x<\pi$. We note that the origin of low-energy spectral weight around $\omega/t=0.1$ at $q_x=0.62\pi$ in Fig.~\ref{fig4}(f) is unclear. Since there is no low-energy structure at corresponding momentum in the $8\times 8$ lattice with square geometry (not shown here), one of possible origins would be a geometry dependent finite-size effect.

\section{Directional dependence of spin excitation}
\label{Sec4}

Recent RIXS experiments have shown the doping dependence of paramagnon excitation~\cite{Meyers2017,Robarts2019,Peng2018} and the difference of excitation energies along the $(0,0)$-$(\pi,0)$ and $(0,0)$-$(\pi,\pi)$ directions~\cite{Meyers2017,Guarise2014,Robarts2019,Monney2016,Peng2018}. In order to check whether the $t$-$t'$-$J$ model can explain this directional dependence or not, it is necessary to examine a system with square geometry having the same boundary condition for both $x$ and $y$ directions. We therefore use an $8\times 8$ $t$-$t'$-$J$ lattice with open boundary condition. We note that the $8\times 8$ lattice is insufficient for describing incommensurate spin excitation near $\mathbf{q}=(\pi,\pi)$ because of only three momenta defined in $0.5\pi<q_x<\pi$.

\begin{figure}[tb]
\center{
\includegraphics[width=0.45\textwidth]{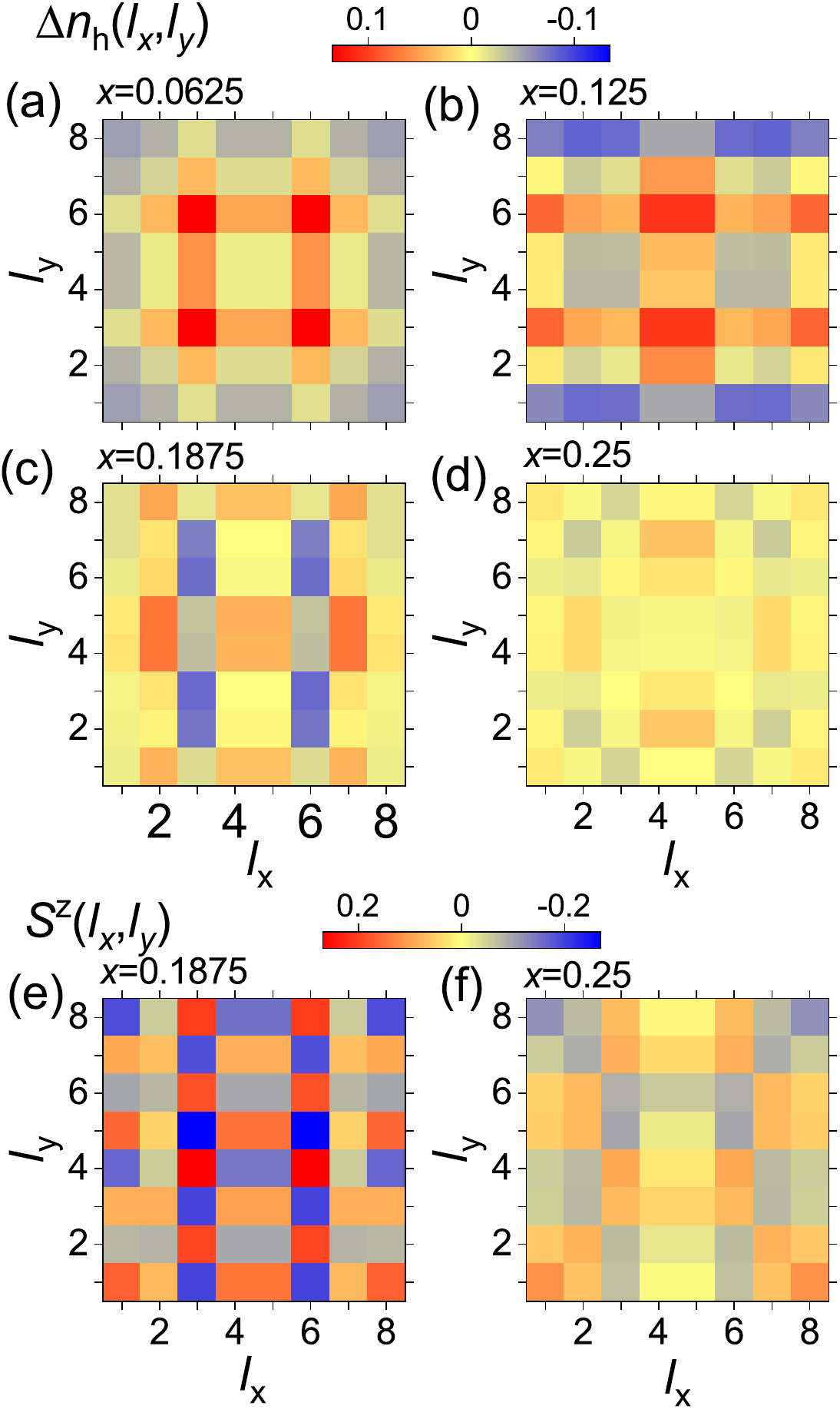}
}
\caption{(Color online) Site-dependent hole density deviated from average valuer $x$, $\Delta n_\mathrm{h}(l_x,l_y)$, for the $8\times 8$ $t$-$t'$-$J$ lattice with open boundary condition ($J/t=0.4$ and $t'/t=-0.25$); (a) $x=0.0625$, (b) $x=0.125$, (c) $x=0.1875$, and (d) $x=0.25$. Each square represents each site and intensity is colored as shown on the top of (b).  Site-dependent spin density $S^z(l_x,l_y)$; (e) $x=0.1875$ and (f) $x=0.25$. Their intensity is colored as shown on the top of (f).}
\label{fig5}
\end{figure}

Before going to spin excitation, we need to clarify the ground state of the $8\times 8$ lattice within our DMRG calculations. Figures~\ref{fig5}(a)-\ref{fig5}(d) exhibit the deviation of hole number at each site from the average value $x$, which is defined by $\Delta n_\mathrm{h}(l_x,l_y)=1-\left<0\right|n_\mathbf{l}\left|0\right>-x$. At $x=4/64=0.0625$ [Fig.~\ref{fig5}(a)], hole carriers tend to be localized at central region with ordered distribution. This may be partly due to the effect of open boundary condition that pushes holes from the boundaries to the center in order to gain kinetic energy. At $x=8/64=0.125$ [Fig.~\ref{fig5}(b)], $\Delta n_\mathrm{h}(l_x,l_y)$ exhibits directional distribution similar to charge stripe, where there are two hole-dominant lines along the $x$ direction at $l_y=3$ and $l_y=6$. In this case, the $x$ direction is chosen as the direction of hole river in our DMRG procedure and we call such a charge distribution the $x$-directional hole river. We may expect that this ground state is degenerate with the state with the $y$-directional hole river that can be obtained if one uses another type of snake connection running to the $x$ direction in our DMRG. At $x=12/64=0.1875$ [Fig.~\ref{fig5}(c)], a stripe-like charge distribution with the $y$-directional hole rivers at $l_x=2$, 4, 5, and 7 appears in the present snake-type DMRG calculation, but it weakens at $x=16/64=0.25$ as shown in Fig.~\ref{fig5}(d).

The spin density $S^z(l_x,l_y)\equiv \left< 0\left| S^z_\mathbf{l}\right| 0\right>$ at each site is very small ($\left|S^z(l_x,l_y)\right| <1.5\times 10^{-4}$) for $x=0.0625$ and $x=0.125$, which means that the spin-reversal symmetry in the system with the $z$ component of total spin being zero is almost kept in the DMRG calculations. On the other hand, $S^z(l_x,l_y)$ is finite for $x=0.1875$ and $x=0.25$ as shown in Figs.~\ref{fig5}(e) and \ref{fig5}(f), respectively. Thus, our DMRG processes for these $x$ pick up only one of spin orientations in contrast to the cases of $x=0.0625$ and $x=0.125$ where a reversed spin orientation is also fully included in the ground state. This is due to insufficient number of $M$ in our calculations. If we were able to increase $M$, for example, the double of the present $M$, we could expect that the reversed spin orientation may equally contribute to the ground state for $x=0.1875$ and $x=0.25$ and result in vanishing $S^z(l_x,l_y)$ as in the case for $x=0.0625$ and $x=0.125$. This remains a future problem. We discuss below the calculated results of $S^z(l_x,l_y)$, assuming that magnetic properties for $x=0.1875$ and $x=0.25$ can be described by the present ground-state calculation where one of spin orientations is expected to be properly taken into account. At $x=0.1875$, we find an antiferromagnetic spin arrangement along the $y$ direction that is the same direction with the $y$-directional hole river seen in Fig.~\ref{fig5}(c), while NN ferromagnetic spin arrangement is observed along the $x$ direction perpendicular to the hole rivers. $S^z(l_x,l_y)$ at $x=0.25$ becomes smaller and shows a $2\times 2$ block spin structure mainly along the edge of the lattice, which will contribute to spin excitation as will be discussed below. 

\begin{figure}[tb]
\center{
\includegraphics[width=0.4\textwidth]{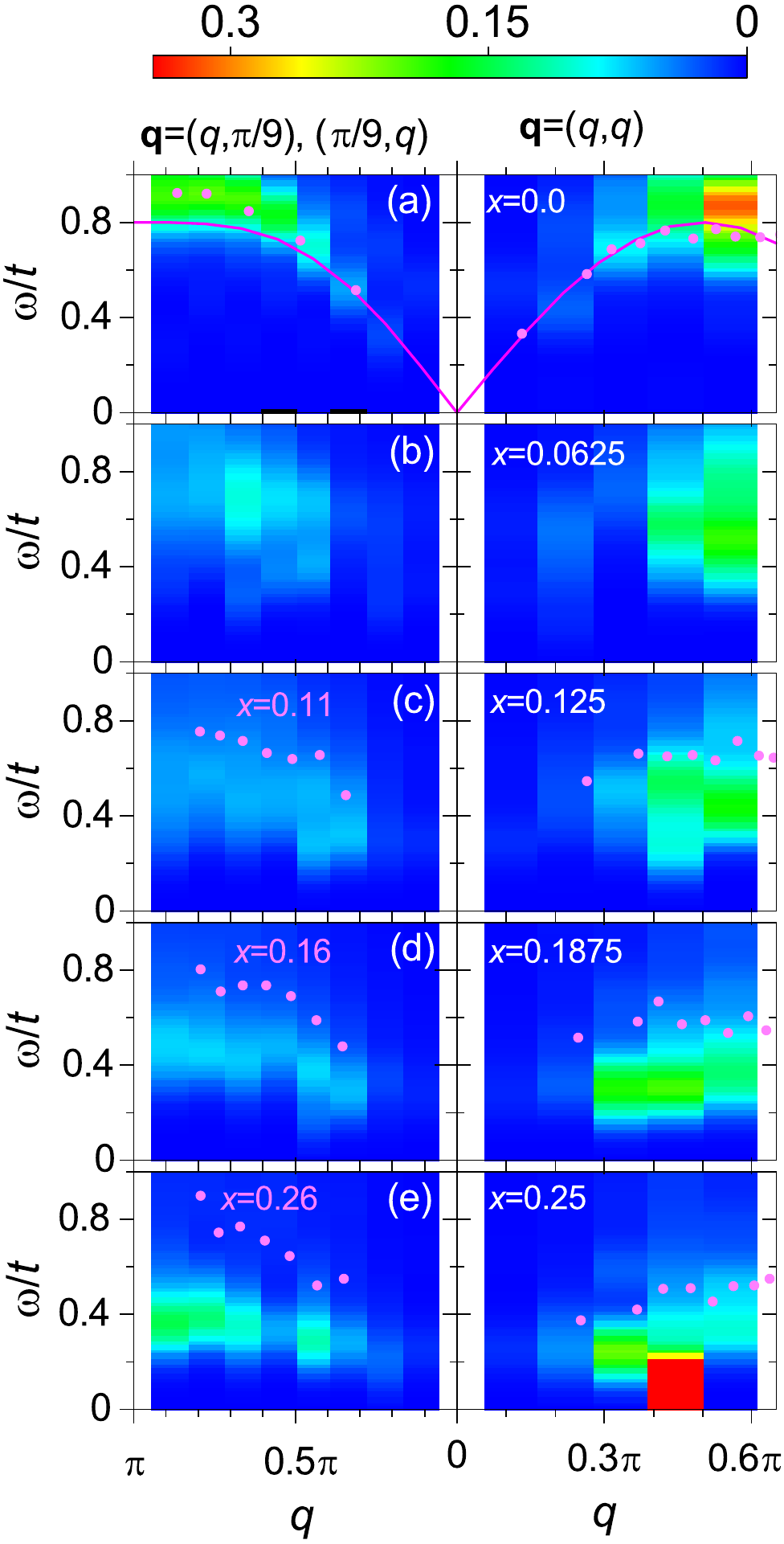}
}
\caption{(Color online) The $x$ dependence of $S(\mathbf{q},\omega)$ from $\mathbf{q}=(8\pi/9,\pi/9)$ and $(\pi/9,8\pi/9)$ to $(\pi/9,\pi/9)$ (left panels) and from $\mathbf{q}=(\pi/9,\pi/9)$ to $(5\pi/9,5\pi/9)$ (right panels) in the $8\times 8$ $t$-$t'$-$J$ lattice with open boundary condition ($J/t=0.4$ and $t'/t=-0.25$). The dots in (a), (c), (d), and (e) represent experimental peak positions of spin excitation in La$_{2-x}$Sr$_x$CuO$_4$ taken from RIXS data~\cite{Meyers2017} for $x=0$, $x=0.11$, $x=0.16$, and $x=0.26$, respectively. For plotting the experimental data on each panel, we assume $t=0.37$~eV. The purple lines in (a) represent spin-wave dispersion at half filling obtained by the linear spin-wave theory for the 2D Heisenberg model. The intensity above 0.35 at $q=0.44\pi$ in the right panel of (e) is colored in red and the maximum intensity is 1.47 at $\omega/t=0.06$.}
\label{fig6}
\end{figure}

Figure~\ref{fig6} shows the doping dependence of $S(\mathbf{q},\omega)$ from $\mathbf{q}=(8\pi/9,\pi/9)$ and $(\pi/9,8\pi/9)$ to $(\pi/9,\pi/9)$ (left panels) and from $\mathbf{q}=(\pi/9,\pi/9)$ to $(5\pi/9,5\pi/9)$ (right panels) in the $8\times 8$ $t$-$t'$-$J$ lattice. For the $(\pi/9,\pi/9)$-$(8\pi/9,\pi/9)$ and $(\pi/9,\pi/9)$-$(\pi/9,8\pi/9)$ directions, averaged spectral weight $[S((q,\pi/9),\omega)+S((\pi/9,q),\omega)]/2$ is plotted. At $x=0$, the lowest-energy excitation follows spin-wave dispersion as seen in Fig.~\ref{fig6}(a) and spectral weights are distributed in the same energy region in both directions. In other words, spin excitation at $\mathbf{q}=(8\pi/9,\pi/9)$ and $(\pi/9,8\pi/9)$ has almost the same energy as the excitation energy at $\mathbf{q}=(4\pi/9,4\pi/9)$. We also plot in Fig.~\ref{fig6}(a) experimental peak positions of spin excitation for La$_2$CuO$_4$ taken from RIXS data along the $(q,0)$ and $(q,q)$ directions~\cite{Meyers2017}, assuming that the calculated peak position at $\mathbf{q}=(7\pi/9,\pi/9)$ and the experimental peak position at $\mathbf{q}=(0.77\pi,0)$ agree each other. This leads to $t=0.37$~eV that is a reasonable value for cuprates. We note that the maximum energy of the experimental spin excitation along the $(q,q)$ direction is slightly lower than that along the $(q,0)$ [$(0,q)$] direction, which is different from our calculated results and the spin-wave dispersion. This is attributed to the effect of cyclic exchange interaction related to four Cu sites~\cite{Coldea2001}, which is not included in our theoretical model.

We first discus our calculated results on the effects of carrier doping in $S(\mathbf{q},\omega)$. With increasing $x$, their intensity reduces accompanied by the shift of spectral weight toward lower energy, which is a common feature of the $t$-$J$-type model resulting from effective reduction of exchange interaction with increasing mobile carriers. As mentioned above, the energy of spin excitation at $\mathbf{q}=(8\pi/9,\pi/9)$ [$(\pi/9,8\pi/9)$] and at $\mathbf{q}=(4\pi/9,4\pi/9)$ is almost the same for $x=0$. In the overdoped region ($x>0.15$), however, the difference of the energy of spin excitation becomes clear. At $x=0.1875$, the main spectral weight at $\mathbf{q}=(8\pi/9,\pi/9)$ and $(\pi/9,8\pi/9)$ is located around $\omega\sim 0.5t$, while the peak position at $\mathbf{q}=(4\pi/9,4\pi/9)$ is at $\omega\sim 0.3t$. Furthermore, low-energy strong spin excitation at $\omega=0.06t$ emerges at $\mathbf{q}=(4\pi/9,4\pi/9)$ for $x=0.25$. This calculated doping dependence indicates that the spin excitation along the $(q,q)$ direction has a tendency toward softening that is stronger than that along the $(p,0)$ [$(0,p)$] directions. Here, we note that the strong low-energy intensity at $\mathbf{q}=(4\pi/9,4\pi/9)$ for $x=0.25$ may partly be related to the $2\times 2$ block-type spin arrangement in Fig.~\ref{fig5}(f), since the wave vector for the block structure is similar to $\mathbf{q}=(4\pi/9,4\pi/9)$ in the momentum space. We also note that low-energy spin excitation near $\mathbf{q}=(\pi/2,\pi/2)$ is observed by a RPA calculation of spin susceptibility for the $t$-$t'$-$U$ Hubbard model~\cite{Guarise2014,Monney2016}, indicating a close connection with spin excitation of our $t$-$t'$-$J$ model in the overdoped region.

Now let us compare our calculated spin excitation with experimental data for La$_{2-x}$Sr$_2$CuO$_4$ taken from the peak positions of RIXS spectra~\cite{Meyers2017}, which are plotted in Figs.~\ref{fig6}(a), \ref{fig6}(c), \ref{fig6}(d), and \ref{fig6}(e). It is clear that quantitative agreement with the experimental data is poor, in the sense that the energies of main spectral weights are located below the experimental peak positions for finite $x$. For more quantitative comparison, we need to include correlated hopping terms related to three sites~\cite{Jia2014} and/or return to original Hubbard-type models~\cite{Peng2018}. From the doping dependence of the experimental data, we can find that the peak positions decrease in energy with increasing $x$ for the $(q,q)$ direction. The decrease is stronger than that for the $(q,0)$ [$(0,q)$] direction. For qualitative level, the decrease of the peak position along the $(q,q)$ direction shares the same trend with the calculated one that is discussed above. Therefore, we may conclude that the calculated doping dependence of spin excitation using $t$-$t'$-$J$ model gives a qualitatively consistent behavior with experimental data, although quantitative agreement is far from complete.

\section{Summary}
\label{Sec5}
		In summary, using dynamical DMRG, we have investigated the dynamical spin structure factor $S(\mathbf{q},\omega)$ in the $m\times n$ $t$-$t'$-$J$ lattice keeping $m\times n=96$ sites. With changing four-leg ladder geometry $24\times 4$ to rectangle geometry $12\times 8$, we found that strong outward dispersion from the incommensurate position toward $\mathbf{q}=(0,\pi)$, which has been reported before for the $24\times 4$ lattice~\cite{Tohyama2018}, looses its intensity followed by the decrease of excitation energy near $\mathbf{q}=(\pi,\pi)$ at hole concentration $x=0.125$. This leads to spectral behaviors consistent with INS data for cuprate superconductors~\cite{Fujita2012}. At the same time, strong incommensurate spin correlation in the $24\times 4$ is reduced in the $12\times 8$ lattice but antiferromagnetic short range correlation remains strong. However, it is interesting to notice that charge stripe with four-lattice period emerges even in the  $12\times 8$ lattice at $x=0.125$ as in the case of the $24\times 4$ lattice. This implies that, while striped charge order induces spin correlation with double period of charge in the $24\times 4$ lattice, such a mechanism does not appear in the $12\times 8$ lattice. Therefore, stripe-driven spin structure is not necessary for understanding spin excitation in INS, but short-range antiferromagnetic correlation is crucial even in the presence of charge stripe.

For a fully squared system, we have examined the $8\times 8$ $t$-$t'$-$J$ lattice with open boundary condition. Even in this system, hole carriers around $x=0.125$ exhibit stripe-type charge distribution in the ground state. Examining the dependence of spin excitation along the $(0,0)$-$(\pi,0)$ and $(0,0)$-$(\pi,\pi)$ directions, we found a softening of spin-excitation energy stronger along the $(0,0)$-$(\pi,\pi)$ direction than along the $(0,0)$-$(\pi,0)$ direction at the overdoped region. This is qualitatively consistent with recent RIXS data~\cite{Meyers2017,Robarts2019}, although the decrease of spin excitation energy with hole doping is stronger than observed one. To make more quantitative comparison, we may need to use original Hubbard-type models, which remains to be a future problem.

\begin{acknowledgment}
We thank M. Mori and M. Fujita for fruitful discussions. This work was supported by MEXT, Japan, as a social and scientific priority issue (creation of new functional devices and high-performance materials to support next-generation industries) to be tackled by using a post-K computer, by MEXT HPCI Strategic Programs for Innovative Research (SPIRE; hp190023), and by the interuniversity cooperative research program of IMR, Tohoku University. The numerical calculation was carried out at the K Computer and HOKUSAI, RIKEN Advanced Institute for Computational Science, and the facilities of the Supercomputer Center, Institute for Solid State Physics, University of Tokyo. This work was also supported by the Japan Society for the Promotion of Science, KAKENHI (Grants No. 17K14148, No. 19H01829, No. JP18H01183, and JP19H05825).
\end{acknowledgment}



\begin{thebibliography}{99}
\bibitem{Fujita2012}For a recent review, see M. Fujita, H. Hiraka, M. Mstsuda, M. Matsuura, J. M. Tranquada, S. Wakimoto, G. Xu, and K. Yamada, J. Phys. Soc. Jpn. {\bf 81}, 011007 (2012) and references therein. 
\bibitem{Tranquada1995}J. M. Tranquada, B. J. Sternlieb, J. D. Axe, Y. Nakamura, and S. Uchida, Nature (London) {\bf 375}, 561 (1995). 

\bibitem{Kaneshita2001}E. Kaneshita, M. Ichioka, and K. Machida, J. Phys. Soc. Jpn. {\bf 70}, 866 (2001). 
\bibitem{Seibold2005}G. Seibold and J. Lorenzana, Phys. Rev. Lett. {\bf 94}, 107006 (2005).
\bibitem{Seibold2006}G. Seibold and J. Lorenzana, Phys. Rev. B {\bf 73}, 144515 (2006).
\bibitem{Kruger2003}F. Kr$\ddot{\mathrm{u}}$ger and S. Scheidl, Phys. Rev. B {\bf 67}, 134512 (2003). 
\bibitem{Carlson2004}E. W. Carlson, D. X. Yao, and D. K. Campbell, Phys. Rev. B {\bf 70}, 064505 (2004). 

\bibitem{Huang2017}E. W. Huang, C. B. Mendl, S. Liu, S. Johnston, H.-C. Jiang, B. Moritz, and T. P. Devereaux, Science {\bf 358}, 1161 (2017).
\bibitem{Huang2017b}E. W. Huang, C. B. Mendl, H.-C. Jiang, B. Moritz, and T. P. Devereaux, npj Quantum Materials {\bf 3}, 22 (2018).

\bibitem{Tohyama1999}T. Tohyama, C. Gazza, C. T. Shih, Y. C. Chen, T. K. Lee, S. Maekawa, and E. Dagotto, Phys. Rev. B {\bf 59}, R11649 (1999). 
\bibitem{White1999}S. R. White and D. J. Scalapino, Phys. Rev. B {\bf 60}, R753 (1999). 
\bibitem{Scalapino2012}D. J. Scalapino and S. R. White, Physica C {\bf 481}, 146 (2012). 
\bibitem{Dodaroo2017}J. F. Dodaro, H.-C. Jiang, and S. A. Kivelson, Phys. Rev. B {\bf 95}, 155116 (2017). 
\bibitem{Tohyama2018}T. Tohyama, M. Mori, and S. Sota, Phys. Rev. B {\bf 97}, 235137 (2018).

\bibitem{Meyers2017}D. Meyers, H. Miao, A. C. Walters, V. Bisogni, R. S. Springell, M. d'Astuto, M. Dantz, J. Pelliciari, H. Y. Huang, J. Okamoto, D. J. Huang, J. P. Hill, X. He, I. Bo\v{z}ovi\'{c}, T. Schmitt, and M. P. M. Dean
Phys. Rev. B {\bf 95}, 075139 (2017). 
\bibitem{Robarts2019}H. C. Robarts, M. Barth\'{e}lemy, K. Kummer, M. Garc\'{i}a-Fern\"{a}ndez, J. Li, A. Nag, A. C. Walters, K. J. Zhou, and S. M. Hayden, Phys. Rev. B {\bf 100}, 214510 (2019). 
\bibitem{Guarise2014}M. Guarise, B. Dalla Piazza, H. Berger, E. Giannini, T. Schmitt, H. M. R{\o}nnow, G. A. Sawatzky, J. van den Brink, D. Altenfeld, I. Eremin, and M. Grioni, Nat. Commun. {\bf 5}, 5760 (2014). 
\bibitem{Monney2016}C. Monney, T. Schmitt, C. E. Matt, J. Mesot, V. N. Strocov, O. J. Lipscombe, S. M. Hayden, and J. Chang, Phys. Rev. B {\bf 93}, 075103 (2016).  
\bibitem{Peng2018}Y. Y. Peng, E. W. Huang, R. Fumagalli, M. Minola, Y. Wang, X. Sun, Y. Ding, K. Kummer, X. J. Zhou, N. B. Brookes, B. Moritz, L. Braicovich, T. P. Devereaux, and G. Ghiringhelli, Phys. Rev. B {\bf 98}, 144507 (2018). 

\bibitem{Sota2010}S. Sota and T.Tohyama, Phys. Rev. B {\bf 82}, 195130 (2010).

\bibitem{Yamada1998}K. Yamada, C. H. Lee, K. Kurahashi, J. Wada, S. Wakimoto, S. Ueki, H. Kimura, Y. Endoh, S. Hosoya, G. Shirane, R. J. Birgeneau, M. Greven, M. A. Kastner, and Y. J. Kim, Phys. Rev. B {\bf 57}, 6165 (1998). 

\bibitem{Sato2014}K. Sato, M. Matsuura, M. Fujita, R. Kajimoto, Sungdae, Ji, K. Ikeuchi, M. Nakamura, Y. Inamura, M. Arai, M. Enoki, and K. Yamada, JPS Conf. Proc. {\bf 3}, 017010 (2014).
\bibitem{Fujita_private}M. Fujita, private communication.

\bibitem{Ido2018}K. Ido, T. Ohgoe, M. Imada, Phys. Rev. B {\bf 97}, 045138 (2018).

\bibitem{Yamase2001}H. Yamase and H. Kohno, J. Phys. Soc. Jpn. {\bf 70}, 2733 (2001).

\bibitem{Coldea2001}R. Coldea, S. M. Hayden, G. Aeppli, T. G. Perring, C. D. Frost, T. E. Mason, S.-W. Cheong, and Z. Fisk, Phys. Rev. Lett. {\bf 86}, 5377 (2001).

\bibitem{Jia2014} C. J. Jia, E.A. Nowadnick, K. Wohlfeld, Y.F. Kung, C.-C. Chen, S. Johnston, T. Tohyama, B. Moritz, and T.P. Devereaux, Nat. Commun. {\bf 5}, 3314 (2014).

\end{thebibliography}
\end{document}